\documentclass[aps,pre,reprint,superscriptaddress,nofootinbib]{revtex4-2}
\usepackage[T1]{fontenc}
\usepackage[utf8]{inputenc} 
\usepackage{xcolor}
\usepackage[unicode=true,pdfusetitle,
bookmarks=true,bookmarksnumbered=false,bookmarksopen=false, breaklinks=true, ,backref=false,colorlinks=true]
 {hyperref}
\hypersetup{citecolor=blue}
\usepackage{babel}
\usepackage{amsmath}
\usepackage{graphicx}

\usepackage{newfloat}
\DeclareFloatingEnvironment[name={Supplementary Figure}]{suppfigure}

\begin{document}
\title{A behavior-environment information loop drives sensory navigation}

\author{Kevin S. Chen}
\email{kevin.s.chen@yale.edu}
\affiliation{Department of Molecular, Cellular, and Developmental Biology, Yale University, New Haven, CT 06511, USA}
\affiliation{Quantitative Biology Institute, Yale University, New Haven, CT 06511, USA}

\author{Matthew P. Leighton}
\affiliation{Quantitative Biology Institute, Yale University, New Haven, CT 06511, USA}
\affiliation{Department of Physics, Yale University, New Haven, CT 06511, USA}

\author{Damon A. Clark}
\email{damon.clark@yale.edu}
\affiliation{Department of Molecular, Cellular, and Developmental Biology, Yale University, New Haven, CT 06511, USA}
\affiliation{Quantitative Biology Institute, Yale University, New Haven, CT 06511, USA}
\affiliation{Department of Physics, Yale University, New Haven, CT 06511, USA}
\affiliation{Department of Neuroscience, Yale University, New Haven, CT 06511, USA}
\affiliation{Wu-Tsai Institute, Yale University, New Haven, CT 06511, USA}

\author{Thierry Emonet}
\email{thierry.emonet@yale.edu}
\affiliation{Department of Molecular, Cellular, and Developmental Biology, Yale University, New Haven, CT 06511, USA}
\affiliation{Quantitative Biology Institute, Yale University, New Haven, CT 06511, USA}
\affiliation{Department of Physics, Yale University, New Haven, CT 06511, USA}
\affiliation{Wu-Tsai Institute, Yale University, New Haven, CT 06511, USA}

\date{\today}

\begin{abstract}

As organisms navigate the environment to locate critical resources, their behavioral actions must be tightly coupled to their sensory inputs.
Here, we introduce an information-theoretic framework that quantifies this coupling using transfer entropy, which measures information flow between sensory inputs and behavioral outputs. Information flow from sensory inputs to behavior defines a ``reactive'' component of a navigational strategy, whereas information flow from behavior to sensory inputs defines an ``active'' component, whereby actions shape subsequent sensory experiences. 
Analyzing these bidirectional information flows enables us to both \emph{predict} navigational performance and \emph{dissect} navigation strategies from trajectories. 
Using a minimal model that captures the active and reactive components, we connect macroscopic performance to microscopic information flows.  
We then apply the framework to experimentally measured trajectories of bacteria, worms, and flies, as well as to machine learning agents navigating sensory landscapes. Across systems, bidirectional information flow reliably predicts navigation efficiency, revealing a common behavioral-environment feedback loop. 
Decomposing active and reactive information flows further exposes distinct strategies underlying bacterial chemotaxis, the spatial dependency of the navigation strategy in fly olfactory navigation, and the learned policies of a reinforcement-trained agent.
Together, these results establish bidirectional information flow as a unifying principle for understanding navigation in biological and artificial systems.


\end{abstract}

\maketitle

\section{Introduction}

Living systems navigate by extracting information from sensory cues and combining it with past experience to decide where to move next \cite{Baker2018-ys, Mobbs2018-au}. 
Navigation involves two complementary flows of information. One runs from sensory signals to the navigator's actions and is commonly described as \textit{reactive}. The other runs from actions back to sensory signals, capturing the fact that an organism’s movements \textit{actively} shape the information it subsequently receives. For instance, bacteria reactively reduce their tumbling rate if they detect that they are moving up a gradient of attractant \cite{Berg1993-yq, Waite2018-lu}. At the same time, tumbling actively alters the signals they will encounter next~\cite{long2017feedback}. Likewise, flies sense intermittent odor encounters to modulate turning and stopping probability \cite{Demir2020-rk, Alvarez-Salvado2018-vl}, while these actions influence the likelihood of future encounters with the odor plume. 

Past work on sensory coding \cite{Sharpee2006-fg,Palmer2015-pz}, behavioral decisions \cite{Gepner2015-wm, Gomez-Marin2011-ok, Izquierdo2015-df}, and memory in navigation \cite{Emonet2024-qy, Siliciano2025-zy, Kathman2024-fr} has largely focused on understanding how information flows from sensory signal to behavioral actions \cite{Gepner2015-wm, Berg1993-yq, Chen2023-fy, Demir2020-rk,Alvarez-Salvado2018-vl}. 
Advances in experimental techniques now enable measurements of freely-moving organisms and decision-making in closed-loop environments, creating an opportunity to study the reciprocal flow of information whereby actions shape future sensory inputs.
While a few studies have explored this feedback during navigation~\cite{long2017feedback,dufour2014limits, Vergassola2007-yb}, a principled framework for quantifying its contribution to navigation and connecting to experimental measurements remains lacking.


Acquiring information about the sensory environment is a natural objective of navigation. Previous work has characterized and optimized navigation using measures such as uncertainty reduction \cite{Vergassola2007-yb, Masson2013-kc}, mutual information between location and source \cite{Boie2018-eb}, the predictability of source location from sensory features \cite{Rigolli2022-bp}, and information-processing constraints on decision making and action \cite{tishby2010information}. Furthermore, Bayesian approaches \cite{Heinonen2023-zy, Yang2018-gq, friston2016active} have provided a foundation for active sensing by describing how agents construct beliefs or internal models of their environment. These approaches have provided important insights into information acquisition and decision-making during navigation. However, they generally rely on explicit models of the environment or focus on particular aspects of information processing. A framework for estimating the bidirectional information flow directly from experimentally measured navigational trajectories remains absent.

In this work, we introduce a framework for quantifying bidirectional information flow between sensory signals and behavioral actions during navigation. Using a minimal bipartite model of coupled behavior and environment dynamics, we derive an information-theoretic metric that connects microscopic information flows to macroscopic navigational performance. Applicable directly to experimental trajectories, the resulting metric provides leading-order predictions of navigation efficiency  across experimental measurements spanning species and environments. Beyond predicting performance, the framework decomposes active and reactive information flows, revealing diverse navigation phenotypes, spatially localized strategies, and emergent policies learned through reinforcement learning.

\section{Results \label{sec: results}}

\subsection{Minimal model for navigation}

\begin{figure}
\begin{centering}
\includegraphics[width=\columnwidth]{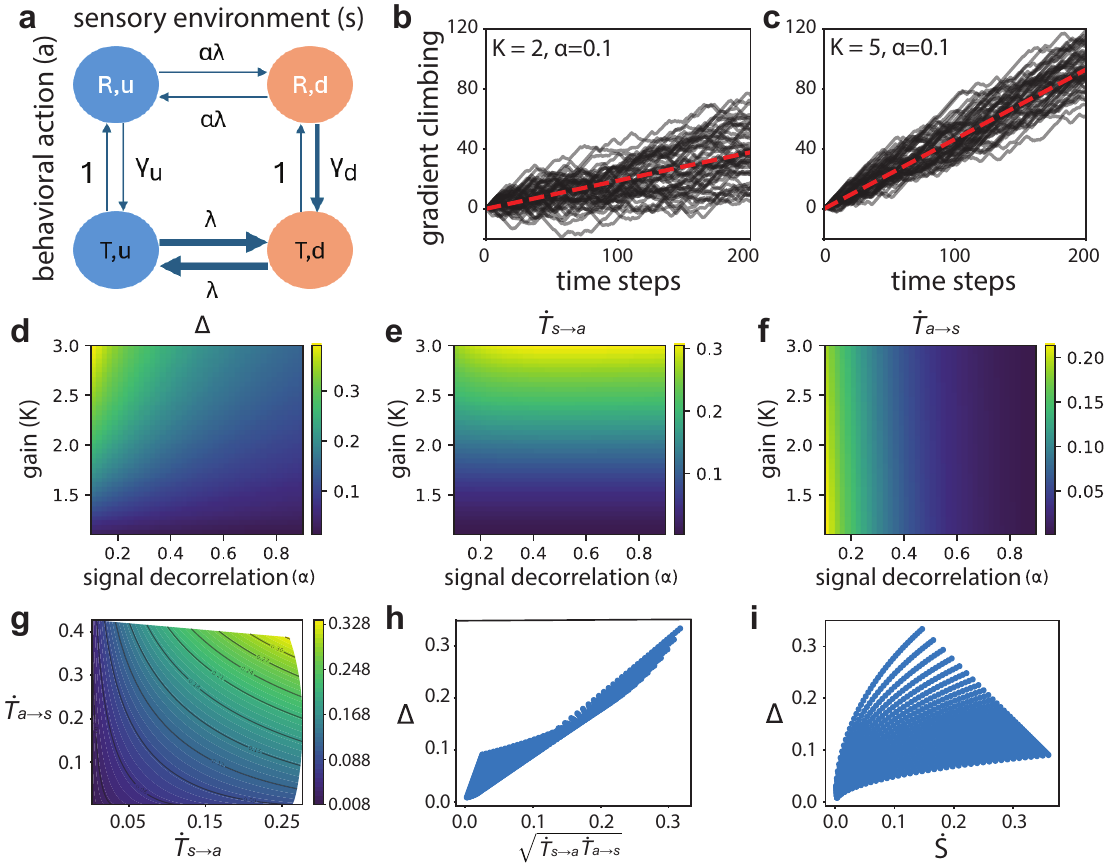}
\par\end{centering}
\caption{Minimal model for navigation reveals the role of bidirectional information flow.  
\textbf{(a)} Bipartite system that describes behavior-environment dynamics.
\textbf{(b)} Simulated navigational trajectories (black tracks) and analytic solution (red dashed line) from the minimal model with $\lambda=1$, $K=2$ and $\alpha=0.1$.
\textbf{(c)} Same as (b) but with the higher gain $K=5$.
\textbf{(d)} Chemotaxis index $\Delta$ as a function of sensory gain $K$ and signal decorrelation factor $\alpha$.
\textbf{(e,f)} Same as (d) but for the two information flows $\dot{T}_{s \rightarrow a}$ and $\dot{T}_{a\rightarrow s}$.
\textbf{(g)} Chemotaxis index as a function of the reactive and active information flows $\dot{T}_{s \rightarrow a}$ and $\dot{T}_{a\rightarrow s}$.
\textbf{(h)} Scatter plot of chemotaxis index as a function of the geometric mean of bidirectional information flow $\sqrt{\dot{T}_{s\rightarrow a}\dot{T}_{a\rightarrow s}}$.
\textbf{(i)} Chemotaxis index as a function of entropy production rate $\dot S$.
%
\label{fig:minimal_model}}
\end{figure}

To build intuition before analyzing experimental data we use a minimal model of navigation inspired by bacterial chemotaxis that comprises two behavioral actions $a \in \{R,T\}$ for runs or tumbles, and two environmental sensory states $s \in \{u,d \}$ for up or down gradient contexts (Fig.~\ref{fig:minimal_model}a). The four possible states $(R,u), (R,d), (T,u), (T,d)$ form a bipartite system with constant transition rates between states. Continuous-time Markov models on bipartite graphs are a common framework to study thermodynamics, information flow, and molecular states \cite{tohme2026fast, horowitz2014thermodynamics, ehrich2023energy, leighton2024information}.


To model navigation we impose a few constraints on the transition rates between states. During tumbles, the organism rapidly reorients between up and down states at the rate $\lambda$ in both directions. During runs, transitions between up and down states occur at the slower rate $\alpha \lambda$ in both directions, with $0 <\alpha<1$ reflecting the longer persistence of direction during runs compared to tumbles. 
We interpret $\alpha$ as a signal decorrelation factor that sets the rate at which environmental noise and imperfect behavioral control (e.g., rotational diffusion) decorrelate the sensory state during runs relative to tumbles \cite{dufour2014limits, long2017feedback}.

 
Chemotaxis requires $\gamma_u < \gamma_d$ so that runs are longer when going up than when going down the gradient. Finally, tumble to run transitions occur at the same rate $\kappa$ regardless of the sensory state.
For convenience, we rescale all rates in the minimal model by this spontaneous rate, setting $\kappa=1$ (Fig.~\ref{fig:minimal_model}a).
The following main results are not sensitive to this choice of constraints on these rate parameters. 

With these definitions, the transition rate matrix of the system reads:
\begin{equation}\label{eq:Qmatrix}
    Q=\left(\begin{array}{cccc}
-(\alpha \lambda+\gamma_u) & \alpha \lambda & \gamma_u & 0 \\
\alpha \lambda & -(\alpha \lambda+\gamma_d) & 0 & \gamma_d \\
1 & 0 & -(\lambda+1) & \lambda \\
0 & 1 & \lambda & -(\lambda+1)
\end{array}\right)
\end{equation}
We solve for the steady state $\pi Q=0$ and obtain probability row vector $\pi=(\pi_{Ru}, \pi_{Rd}, \pi_{Tu}, \pi_{Td})$ \eqref{eq:pi_steady_state} from which we can calculate metrics of performance and behavior, namely the chemotaxis index $\Delta=\pi_{Ru}-\pi_{Rd}$, which quantifies the average speed up the gradient, and the tumble bias $\text{TB}=\pi_{Tu}+\pi_{Td}$, which quantifies the fraction of time spent tumbling:
\begin{align}
    \Delta & =\frac{\gamma_d-\gamma_u}{Z}\;,
    \label{eq:Delta}
\\
  \text{TB} &=\frac{\alpha\left(1+\lambda\right) \left(\gamma_d+\gamma_u\right)+2\gamma_d\gamma_u}{Z}\;.
  \label{eq:TB}
\end{align}
with the normalization $Z = \alpha\left(\gamma_d+\gamma_u\right)(2 \lambda+1)+2 \alpha(2 \lambda+1)+2 \gamma_d \gamma_u+\gamma_d+\gamma_u$. 

Despite its simplicity, this minimal model can implement a range of navigation strategies. To build intuition we define $\gamma_d=K$ and $\gamma_u=1/K$, with $K>1$, which reduces the control strategy to one parameter sensory gain $K$, whereas $\alpha$ controls how long the signal remains correlated during runs. The chemotaxis performance becomes:
\begin{equation}
    \Delta=\frac{K-1}{(K+1)(2 \alpha \lambda+\alpha+1)}\;.
    \label{eq:delta_K}
\end{equation}
This function increases monotonically with sensory gain $\partial_K \Delta|_{\lambda, \alpha}>0$, and decreases monotonically with signal decorrelation $\partial_\alpha \Delta|_{\lambda, K}<0$,
indicating that signal correlations and high sensitivity lead to good navigation performance. Stochastic simulations are consistent with these predictions over a wide range of values of the sensory gain $K$ and decorrelation factor $\alpha$ (Fig.~\ref{fig:minimal_model}b-d)

In this minimal model, the transition rates between sensory states conditioned on action are symmetric, whereas the transition rates between action states conditioned on sensory states are not. We now show that this asymmetry results in an information loop that flows from signal to action and back, and governs navigational performance.

\subsection{Linking navigational performance to the information loop between signal and behavior}

We consider the coupled time series of sensory inputs experienced by an agent, $s_t$, and the behavioral actions it takes, $a_t$, at time $t$. 
Although the mutual information rate $\dot{I}(s;a)$ characterizes information processing in sensory systems \cite{Sharpee2006-fg, Palmer2015-pz, strong1998entropy}, it is symmetric and therefore does not resolve the direction of information flow. In a bipartite system, the absence of simultaneous transitions between the states of $a$ and $s$ allows the mutual information rate to be decomposed into two directional transfer entropy rates \cite{hartich2017stochastic,mattingly2021escherichia}:
\begin{equation}
    \dot{I}(s;a) = \dot{T}_{s \rightarrow a} + \dot{T}_{a \rightarrow s}\;.
\end{equation}
We refer to $\dot{T}_{s \rightarrow a}$ as the \textit{reactive} information flow, as it quantifies how sensory signals shape future actions, and to $\dot{T}_{a \rightarrow s}$ as the \textit{active} information flow, as it quantifies how actions generate future sensory information. More formally, $\dot{T}_{s \rightarrow a}$ quantifies the reduction in uncertainty about the next action provided by the history of sensory inputs, conditioned on the history of actions~\cite{Schreiber2000-ir, mattingly2021escherichia}, and $\dot{T}_{a \rightarrow s}$ is defined similarly. 

Computing transfer entropy rates from data is generally infeasible because it requires estimating probabilities conditioned on the entire past. We therefore consider the simpler one-step transfer entropy rates 
\begin{equation}\label{eq:transfer_entropy_rate_one_step}
    \dot{T}_{s \rightarrow a} = \lim_{dt \rightarrow 0}   
     \frac{1}{dt}\left[H(a_{t+dt}|a_t) - H(a_{t+dt}|a_t,s_t)\right]\;,
\end{equation}
and similarly for $\dot{T}_{a \rightarrow s}$. Here $H(a_{t+dt}|a_t)$ is the conditional uncertainty or Shannon entropy about the next action $a_{t+dt}$ given the current action $a_t$, and the difference with $H(a_{t+dt}|a_t,s_t)$ quantifies the reduction in uncertainty (information gain) due to knowledge of the current sensory state $s_t$. These one-step transfer entropy rates are upper bounds on the full-history transfer entropy rates~\cite{hartich2017stochastic}. 
We examine history dependence in the minimal model and show that transfer entropy rates computed from the full history are qualitatively consistent with those obtained from one-step calculations (Appendix~\ref{appendix: single vs multi-step}).

For the minimal model, the one-step transfer entropy rates are defined by rate parameters (Appendix~\ref{eq:transfer_entropy_rate_one_step_markov}):
\begin{align}
\dot{T}_{s \rightarrow a} &=
\pi_{R u}\,\gamma_u \log \frac{\gamma_u}{\bar{q}_{R\rightarrow T}}
+
\pi_{R d}\, \gamma_d \log \frac{\gamma_d}{\bar{q}_{R\rightarrow T}} \label{eq:entropy_transfer_rate_s_a}
\\
\dot{T}_{a \rightarrow s} &=
\pi_{Ru}\,\alpha\lambda\ln\frac{\alpha\lambda}{\bar q_{u\to d}}
+
\pi_{Tu}\,\lambda\ln\frac{\lambda}{\bar q_{u\to d}}
\nonumber\\
&\quad+
\pi_{Rd}\,\alpha\lambda\ln\frac{\alpha\lambda}{\bar q_{d\to u}}
+
\pi_{Td}\,\lambda\ln\frac{\lambda}{\bar q_{d\to u}}
\label{eq:entropy_transfer_rate_a_s}
\end{align}
%
%
where $\bar{q}_{R\rightarrow T}=\frac{\pi_{R u} \gamma_u+\pi_{R d} \gamma_d}{\pi_{R u}+\pi_{R d}}$, $\bar q_{u\to d} =\frac{\pi_{Ru}\,\alpha\lambda+\pi_{Tu}\,\lambda}{\pi_{Ru}+\pi_{Tu}}$, and $\bar q_{d\to u}=\frac{\pi_{Rd}\,\alpha\lambda+\pi_{Td}\,\lambda}{\pi_{Rd}+\pi_{Td}}$. The asymmetry between the transition rates mentioned in the previous section is inherited by the transfer entropy rates: in $\dot{T}_{s \rightarrow a}$ only run-to-tumble transitions are informed by the signal, whereas in $\dot{T}_{a \rightarrow s}$, both up-to-down and down-to-up signal transitions are informed by the actions. 

To further examine the consequence of such asymmetry, we take a perturbative approach. 
Living organisms often navigate highly stochastic environments while operating under weak control authority. Therefore, we consider the limits where $\alpha\approx 1$ (noisy environment that results in fast signal decorrelation) and $\gamma_d\approx\gamma_u$ (weak control). From~\eqref{eq:entropy_transfer_rate_s_a}-\eqref{eq:entropy_transfer_rate_a_s} we obtain 
\begin{equation}
    \lim _{\gamma_u \rightarrow \gamma_d} \dot{T}_{s \rightarrow a}=0, \quad \lim _{\alpha \rightarrow 1} \dot{T}_{a \rightarrow s}=0\;.
\end{equation}
When $\gamma_u=\gamma_d$, transitions between action states become independent of the sensory states, eliminating information flow from sensory input to action ($\bar q_{R\rightarrow T}=\gamma$ in \eqref{eq:entropy_transfer_rate_s_a}). Conversely, when $\alpha=1$, environmental noise is such that transitions between sensory states occur at the same rate during runs and tumbles, eliminating information flow from action to signal ($\bar q_{u\rightarrow d} = \bar q_{d\rightarrow u} = \lambda$ in \eqref{eq:entropy_transfer_rate_a_s}). 
%
%
Expanding the transfer entropy rates around these limits, we obtain (see derivation in the Supplementary Information):
\begin{equation}
\dot{T}_{s \rightarrow a}=\frac{1}{8 \gamma_u(\gamma_u+1)}(\gamma_d - \gamma_u)^2+\mathcal{O}\left[(\gamma_d - \gamma_u)^3\right],
\end{equation}
\begin{equation}
\dot{T}_{a \rightarrow s}=\frac{\gamma_u\lambda}{2(\gamma_u+1)^2}(1-\alpha)^2+\mathcal{O}\left[(1-\alpha)^2(\gamma_d-\gamma_u)\right].  
\end{equation}
To leading-order $\dot{T}_{s \rightarrow a}$ scales as $(\gamma_d-\gamma_u)^2$ whereas $\dot{T}_{a \rightarrow s}$ scales as $(1-\alpha)^2$. Thus, 
increasing the asymmetry in the run-to-tumble transition rates enhances information flow from signal to action (Figure~\ref{fig:minimal_model}e), while decreasing $\alpha$, for example by increasing environmental persistence, enhances information flow from action to signal (Figure~\ref{fig:minimal_model}f). Since all transition rates have been rescaled by $\kappa$, the transfer entropy rates are dimensionless. Unless otherwise mentioned for experimental data, we report information rates in bits per unit time for the models.

We saw above that navigational performance increases with  $(\gamma_d-\gamma_u)$ and decreases with $\alpha$, suggesting a direct connection between performance and information flows. To make this connection quantitative, we performed a similar perturbative expansion of the chemotaxis index $\Delta$. Matching orders of the expansions yields the following scaling relationship between navigational performance and transfer entropy  (see derivation in the Supplementary Information) 
\begin{equation}
    \Delta \propto \text{A}\,\sqrt{\dot{T}_{s\rightarrow a}} + \text{B}\,\sqrt{\dot{T}_{s\rightarrow a} \dot{T}_{a\rightarrow s}} - \text{C} \,\dot{T}_{s\rightarrow a} \equiv \Phi
    \label{eqn: Phi}\;,
\end{equation}
where $A=\frac{\sqrt{2\gamma_u/(1+\gamma_u)}}{(1 + 2\lambda+\gamma_u)}$, $B=\frac{2(2\lambda+1)\sqrt{(1+\gamma_u)/\lambda}}{(1 + 2\lambda+\gamma_u)^2}$, $C=\frac{4\gamma_u(1+\lambda+\gamma_u)}{(1+\gamma_u)(1 + 2\lambda+\gamma_u)^2}$ are positive functions of $\lambda$ and $\gamma_u$ but not $\alpha$ or $\gamma_d$.
We refer to $\Phi$ as the leading-order information-theoretic predictor of navigational performance that accounts for bidirectional information flow. 
The first term recovers the previously derived square root dependence of performance on the reactive information flow in shallow gradient \cite{mattingly2021escherichia}, while the second term introduces the contribution of the active information flow through the geometric mean of the two transfer entropy rates. For parameter values typical of the minimal model, the coefficient $B$ is generally the largest and the third term provides smaller corrections. Thus, the coupling between the reactive and active information flows plays a dominant role in determining performance.

This interpretation is supported by the phase diagram of the minimal model. Parameter combinations yielding large chemotaxis indices exhibit comparable reactive and active information flows, rather than large values of either term alone (Fig.~\ref{fig:minimal_model}g). Consequently, successful navigation requires both reactive and active information flows to be substantial, forming an information ``loop'' in which sensory signals guide actions and actions, in turn, generate informative sensory experiences. Consistent with this picture, the geometric mean of the two transfer entropy rates strongly correlates with the chemotaxis index (Fig.~\ref{fig:minimal_model}h, $R^2=0.9$) indicating that the coupling between reactive and active information flows is a powerful predictor of navigation performance (Fig.~\ref{fig:minimal_model}e,f,g).


Alternative quantities are not as correlated with performance (Appendix~\ref{appendix: metrics}): the mutual information between sensory and action states yields $R^2\approx 0.8$, the square root of the reactive information flow alone yields $R^2\approx 0.4$, and the non-equilibrium flux along the bipartite graph yields $R^2\approx 0.1 $, compared with $R^2\approx 0.95$ for the information-theoretic predictor $\Phi$. 
Since directed navigation requires a net probability circulation on the bipartite state graph, we also computed the entropy production rate of the dynamics (Fig.~\ref{fig:minimal_model}i). Although entropy production is positively correlated with navigation performance, the relationship exhibits substantial scatter ($R^2\approx0.2$), indicating that dissipation alone is a poor predictor of performance. This observation is consistent with recent work showing that entropy production is necessary but not sufficient for biological function \cite{tu2018adaptation}.

The central conclusion of this analysis is that reactive and active information flows play fundamentally different roles in navigation, an asymmetry that is encoded directly in the predictor $\Phi$. Because every contribution to $\Phi$ vanishes when
$\dot T_{s\rightarrow a}=0$, navigation becomes impossible in the absence of reactive information flow. 
In contrast, $\dot{T}_{a\rightarrow s}$ enters only through the mixed term $\sqrt{\dot{T}_{s\rightarrow a}\dot{T}_{a\rightarrow s}}$, so navigation remains possible even when the active information flow vanishes. This regime corresponds to a purely reactive strategy, in which the agent biases its motion up the gradient but can no longer exploit environmental correlations to generate sensory information.

Finally, in the Supporting Information we show that the leading-order form of $\Phi$ in \eqref{eqn: Phi} holds when the specific rate constraints of the minimal model are relaxed and even when additional sensory or action states are introduced (Supplementary Information), suggesting that the relationship between navigational performance and bidirectional information flow is robust beyond this minimal model.



\subsection{Bidirectional information flow predicts navigation performance across navigating systems}
\begin{figure}
\begin{centering}
\includegraphics[width=1.\columnwidth]{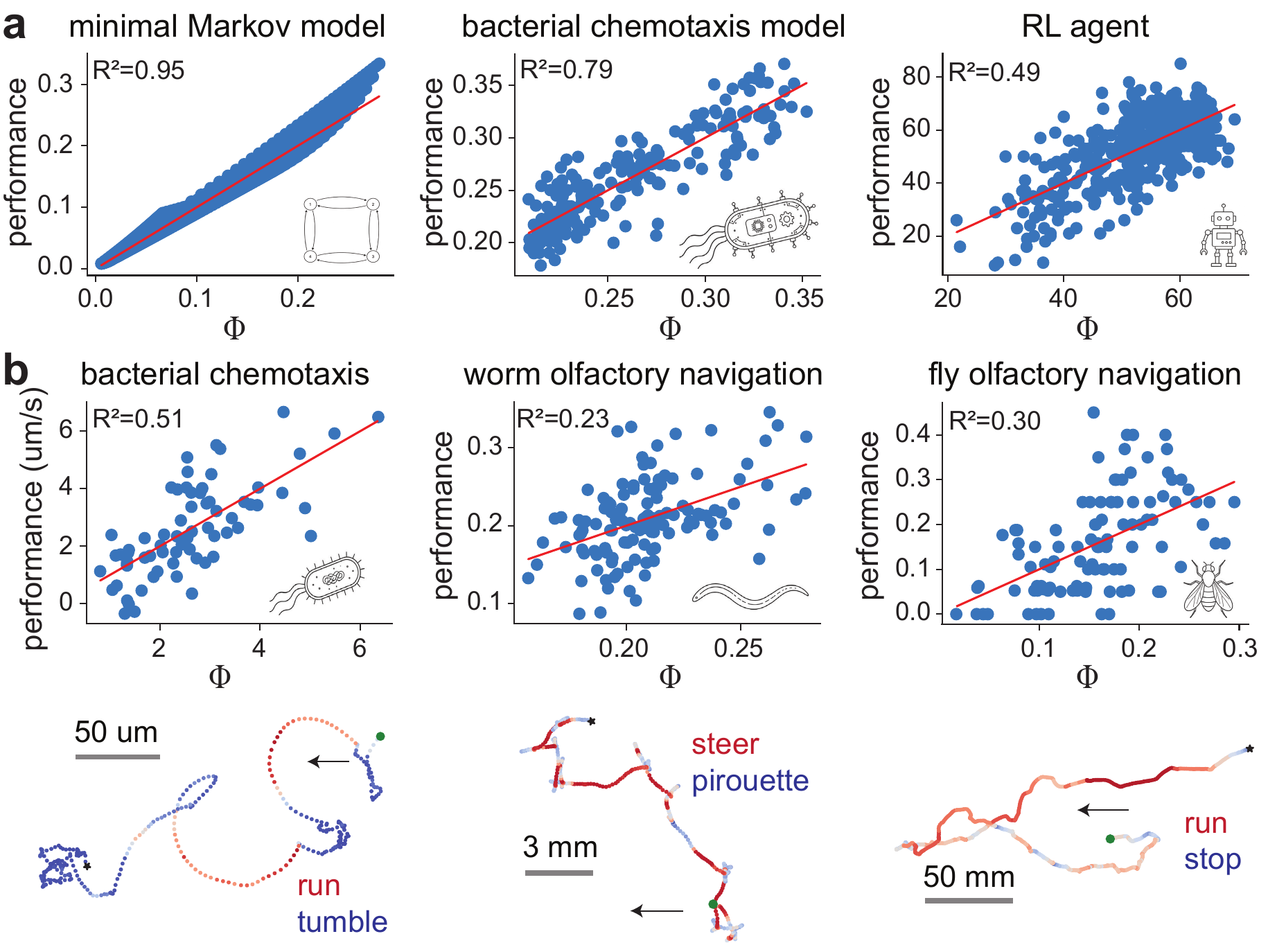}
\par\end{centering}
\caption{Bidirectional information flow between sensing and action predicts performances across navigation data. 
\textbf{(a)} Navigational performance as a function of the bidirectional information metric $\Phi$ across simulated data from computational models. 
\textbf{(b)} Results from measured biological systems, with example navigation trajectories shown below. 
Data contained $>5000$ hour-length of \emph{E. coli} trajectories, $>600$ hour-length of \emph{C. elegans} trajectories, and $>13$ hour-length of \emph{Drosophila} trajectories.
The two behavioral states are shown in color-coded example trajectories. The scale (gray bar), goal-direction (arrow), and initial (green dot) and end points (black star) are indicated.
\label{fig: all data}
}
\end{figure}

To test whether $\Phi$ predicts navigational performance beyond the minimal model, we apply our framework to experimentally measured trajectories of swimming bacteria navigating a linear gradient of attractant \cite{Waite2018-lu}, crawling worms navigating a Gaussian-shaped chemical landscape \cite{Chen2023-fy}, and walking flies navigating an intermittent odor plume that mimics the signal statistics an animal would experience in a turbulent environment~\cite{Demir2020-rk}. 
We include a model for a simplified biophysical model for bacteria chemotaxis \cite{long2017feedback}. We also include a reinforcement learning (RL) agent conducting a navigation task. In all cases, the behavioral measurements consist of trajectories of translational and rotational velocities. The measured sensory signal corresponds to the attractant concentration for the bacteria and worms, the odor encounter events for flies, and the reward signal for the RL agents. 
Performance is computed directly from the raw trajectories in each data set. 
For the chemotaxis model, the performance the normalized drift velocity. 
For RL agent, it is the accumulated reward. 
For bacteria, the performance is the mean of velocity projected onto the gradient direction \cite{mattingly2021escherichia}. For worms, the performance is the normalized difference of concentration change from the initial to final locations \cite{chen2026state}. For flies, the navigation performance is the fraction of trajectories that located the odor source.

To apply our framework to these data, the one-step transfer entropy rates~\eqref{eq:transfer_entropy_rate_one_step} are estimated using the experimental sampling interval and empirical transition probabilities between consecutive states. As a preprocessing step, we coarse grain the measured sensory inputs and behavioral outputs into binary time series of sensory cues and behavioral actions. 
For bacteria, worms, and the RL agent, sensory input is binarized into up- and down-gradient states. For flies, sensory input is binarized into odor-present and odor-absent states by thresholding the intermittent odor intensity signal. To extract discrete action states from continuous behavioral measurements, we construct maximally predictive Markov states following recent work \cite{Costa2024-ho} (Appendix~\ref{appendix: state-space}). Intuitively, these are behavioral states that retain as much predictive information about future behavior as possible. The method partitions the delay-embedding space of the trajectories into discrete states and approximates the resulting transfer operator with a Markov model. Applying this unsupervised procedure to the navigation trajectories reveals discrete states that recover known behavioral strategies across datasets, such as run-and-tumble in the bacteria \emph{E. coli} \cite{Waite2018-lu}, steer-and-pirouette in the worm \emph{C. elegans} \cite{Chen2023-fy}, and walk-and-stop in the fruit fly \emph{Drosophila} \cite{Demir2020-rk}.
Once the data have been converted into binary time series, we estimate the transition probabilities between consecutive sensory and behavioral states from the observed time series. 
These empirical transitions are used to 
directly compute the reactive and active information flows $\dot{T}_{s\rightarrow a}$ and $\dot{T}_{a\rightarrow s}$.
Because individual trajectories typically contain only a few state transitions, we estimate information flows and navigation performance from ensembles of trajectories rather than from single trajectories (Appendix~\ref{appendix: state-space}). Finite-size corrections for information estimates are then applied to each ensemble (Appendix~\ref{appendix: finite-size}).

\begin{figure}
\begin{centering}
\includegraphics[width=1.\columnwidth]{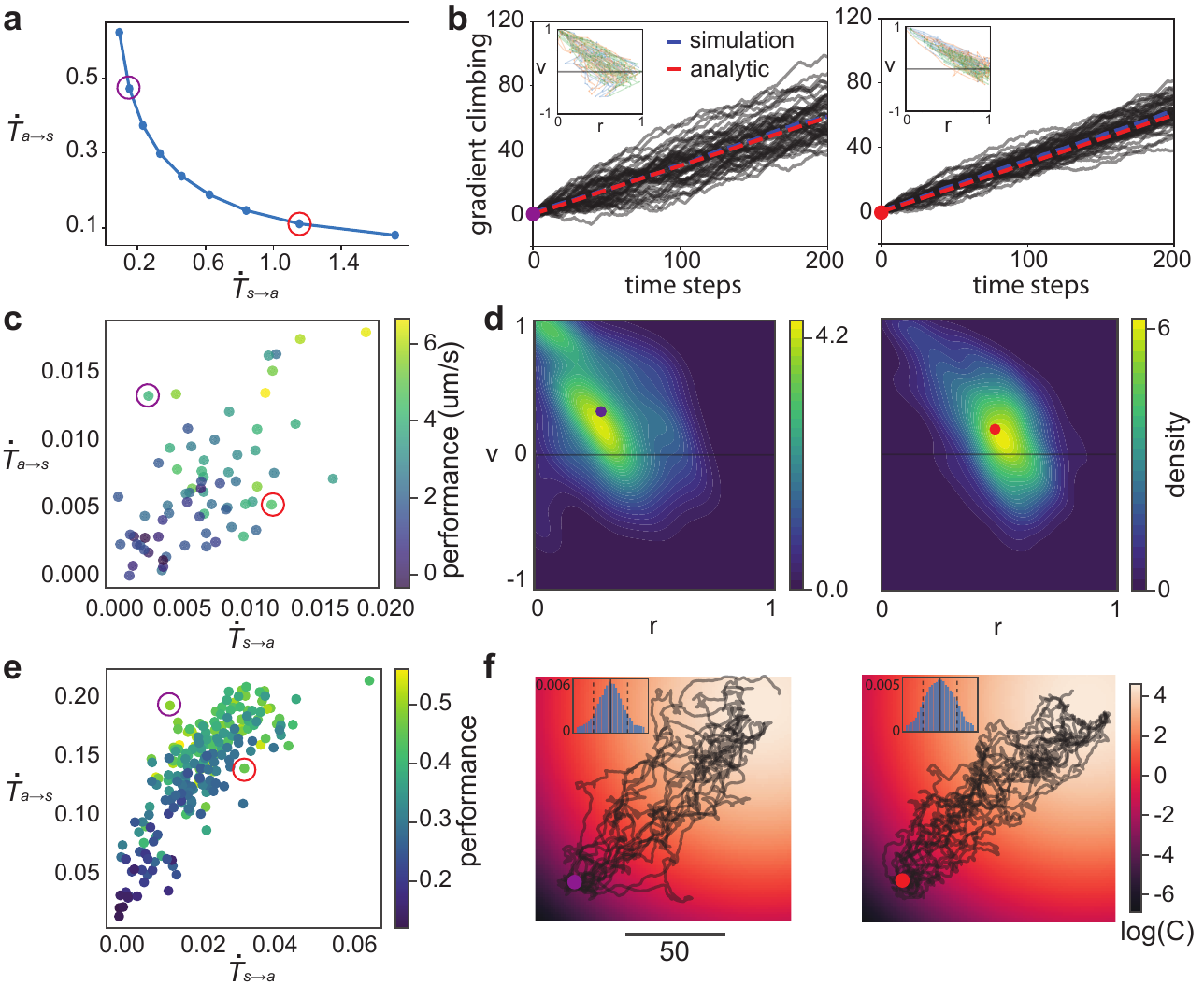}
\par\end{centering}
\caption{Bidirectional information flow reveals diverse strategies in bacteria chemotaxis.  
\textbf{(a)} Iso-performance analysis of the minimal model. An example contour line for performance of the minimal model $\Delta=0.3$ as a function of the reactive and active information flows. 
\textbf{(b)} Left: example trajectories for the purple circle in (a), where $\dot{T}_{a\rightarrow s}$ is larger than $\dot{T}_{s\rightarrow a }$. Right: example trajectories for the red circle in (a). where $\dot{T}_{s\rightarrow a}$ is larger than $\dot{T}_{a\rightarrow s }$. 
Insets: phase portraits of drift $v$ and tumbling probability $r$.
\textbf{(c)} Active and reactive transfer entropy rates (bits/s) estimated from experimental trajectories of \textit{E. coli} (RP437) bacteria navigating a static gradient of methyl-aspartate that ranges from 0 mM to 4 mM over 10 mm. 
Performance is measured through drift velocity in $\mu$m/s. Purple and red circles: example trajectories that are more active or reactive, with drift velocities 0.4 and 0.45 $\mu$m/s, respectively. 
\textbf{(d)} The density of trajectories from the purple (left) and red (right) ensembles in (c) in the phase portraits of $v$ and $r$. Here drift is the probability of running up-gradient and these values are computed in a 5 second sliding time window. Dots show the mean values.
\textbf{(e)} Same as (c) but for simulations of a stochastic biophysical agent-based model of bacterial chemotaxis. 
\textbf{(f)} Example trajectories for purple and red shown in (e), where tracks that are more active or reactive, respectively. The insets show density of angles to the goal direction. Zero degrees is shown as a solid line and $\pm 90$ degrees as dash lines. Scale bar in units of run speed over one characteristic time.
\label{fig: bacteria}}
\end{figure}

Across all systems considered, from minimal and biophysical models to RL agents and experimental trajectories, the bidirectional flow metric $\Phi$ provides a first-order predictor of navigational performance (Fig.~\ref{fig: all data}). The agreement is strongest in the theoretical models (Fig.~\ref{fig: all data}a) and remains significant in experimental datasets spanning bacterial chemotaxis, worm navigation, and fly plume tracking (Fig.~\ref{fig: all data}b), demonstrating that the framework generalizes far beyond the minimal Markov setting. Model comparisons further reveal that the geometric mean of the active and reactive information flows is necessary to account for navigational performance (Table \ref{tab:model_comparison}), underscoring the importance of bidirectional sensorimotor information exchange.



\subsection{Information flow reveals diverse strategies in bacterial chemotaxis}

Because $\Phi$ depends on the geometric mean of the active and reactive information flows,  
navigation exhibits a degeneracy in which different combinations of bidirectional information flow can yield equivalent performance. 

To illustrate this degeneracy, we express the parameters $K$ and $\alpha$ of the minimal model \eqref{eq:delta_K} in terms of the reactive and active information flows and plot an iso-performance contour  (Fig.~\ref{fig: bacteria}a). 
Although all points along this contour have the same navigational performance, they correspond to qualitatively different trajectories. Higher active information flow produces more pronounced ``ratcheting'' dynamics in the phase space of tumble probability and effective drift, whereas strategies dominated by reactive flow exhibit rapid heading correction through increased tumbling following down-gradient motion (Fig.~\ref{fig: bacteria}b). These distinct dynamical regimes are reminiscent of those previously identified in biophysical models of bacterial chemotaxis \cite{long2017feedback}.

We perform a similar analysis on bacterial chemotaxis experimental data by estimating bidirectional information flows from ensembles of trajectories grouped by tumble bias, a phenotype known to be inversely correlated with chemotaxis performance \cite{Waite2018-lu} (Fig.~\ref{fig: bacteria}c). The inferred information flows are on the same order ($\approx 0.01$ bit/s) as those measured directly in the chemotaxis signaling pathway \cite{mattingly2021escherichia}. As expected, tumble bias is negatively correlated with chemotaxis performance. However, we also find ensembles with comparable chemotactic performance but distinct combinations of reactive and active information flows, mirroring the degeneracy and additional dimension of diversity predicted by the minimal model. Furthermore, we can place trajectories dominated by the active and reactive information flows in the phase plane of drift velocity and tumbling probability (Fig.~\ref{fig: bacteria}d). Consistent with the minimal model (Fig.~\ref{fig: bacteria}b, insets), the active-dominated trajectories have lower tumble probability and more aligned runs up-gradient, whereas reactive-dominated trajectories have densities shifted to higher tumble probability.

We next consider a simplified biophysical model of bacterial chemotaxis \cite{long2017feedback}. By sampling a broad range of model parameters, we generate  agents with diverse chemotactic performances and again find that similar performance can arise from distinct combinations of active and reactive information flows (Fig.~\ref{fig: bacteria}e).
We examine simulations from active and reactive-dominated trajectories and find that along the course of navigating to a point source, the active-dominated regime produces narrower heading to the goal direction (Fig.~\ref{fig: bacteria}f, inset).
This result is consistent with previous work identifying gain and persistence as the key dimensions governing behavioral variability in bacterial chemotaxis \cite{long2017feedback}. 
Here, however, the dimensions emerge directly from observed trajectories in a model-free manner and are expressed in terms of interpretable information-theoretic quantities.

Taken together, the minimal model, experimental bacterial data, and the biophysical model reveal a degeneracy of chemotactic strategies: comparable navigation performance can be achieved through different balances of active and reactive information flows.


\subsection{Local plume statistics shape information flow during navigation}

\begin{figure}
\begin{centering}
\includegraphics[width=.72\columnwidth]{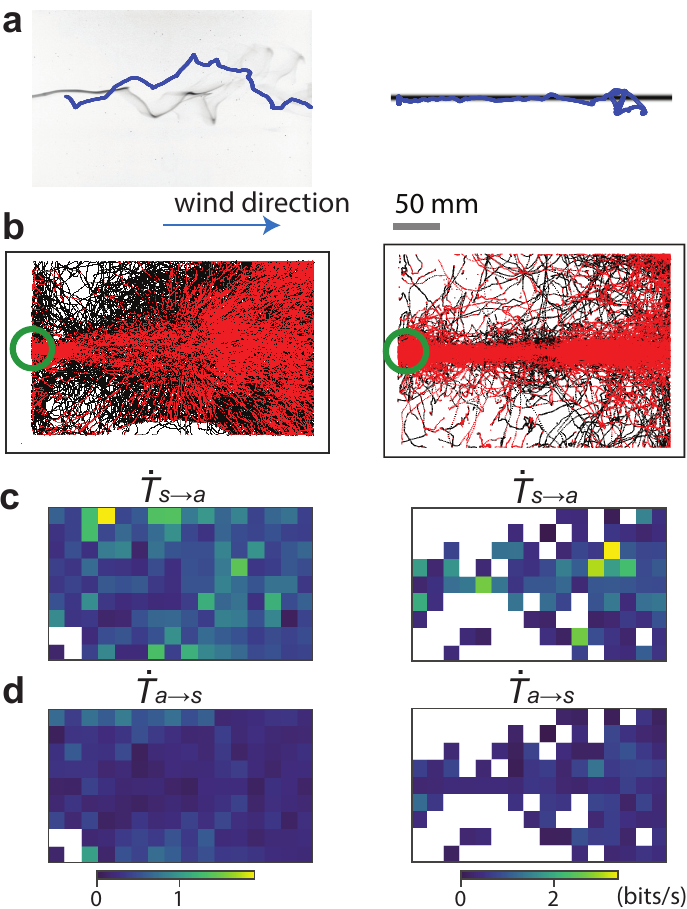}
\par\end{centering}
\caption{Flies navigating complex plumes are more reactive downwind and more active upwind. 
\textbf{(a)} An example navigation trajectory in the complex plume (left) and straight plume (right) environments. The complex plume is time-varying and this is a snapshot of the smoke intensity, whereas the straight plume is stable through time, with dark regions showing higher concentration of smoke. Wind direction and scale are shown in the bottom.
\textbf{(b)} Same as (a) but with all navigation trajectories overlaid. Odor detection events are shown in red and the source location is circled in green.
\textbf{(c)} The reactive information flow $\dot{T}_{s \rightarrow a}$ conditioned on navigation measurements in space. 
\textbf{(d)} Same as (c) but for the active information flow $\dot{T}_{a \rightarrow s}$. Color code is in bits/s and white indicates insufficient data.
\label{fig: fly}}
\end{figure}

Unlike the steady environments of the minimal model and bacterial chemotaxis, many animals navigate complex dynamical olfactory environments with fluctuating spatiotemporal signals~\cite{Celani2014-ki, reddy2022olfactory}. For instance, fruit flies efficiently track intermittent odor plumes~\cite{Demir2020-rk} (Fig.~\ref{fig: fly}a). Understanding how sensory information is used in these environments has motivated both information-theoretic and biologically inspired navigation algorithms \cite{Vergassola2007-yb, Demir2020-rk}. Here, we ask whether bidirectional information flows can reveal how navigation strategies depend on local plume statistics.

To characterize how navigation strategies vary in space, we compute the active and reactive information flows as a function of location within intermittent and ribbon plume environments based on their time-varying odor encountering events ~\cite{Demir2020-rk} (Fig.~\ref{fig: fly}b). In the intermittent plume, reactive information flow is strongest downstream, where odor encounters are highly variable and poorly predictive, whereas active information flow is localized near the plume boundaries upstream, where flies repeatedly enter and exit the plume while orienting toward its center (Fig.~\ref{fig: fly}c,d). 
This is consistent with recent studies showing that flies can actively track edges in olfactory environments \cite{siliciano2025vector}. 
Additionally, we applied the same analysis to fly navigation in a more deterministic environment, which consists of a stable ribbon structure that provides odorants (Fig.~\ref{fig: fly}a, right). 
The ribbon plume shows a similar, though weaker, downstream enhancement of reactive information flow, suggesting that this component partly reflects how flies engage in the navigation task from downwind locations~(Fig.~\ref{fig: fly}c,d; right).

Together, these results show that bidirectional information flows reveal spatially localized  navigation strategies shaped by the statistics of the sensory environment.

\subsection{Bidirectional information flow reveals operational regimes of a two-armed bandit task}

\begin{figure}
\begin{centering}
\includegraphics[width=1.\columnwidth]{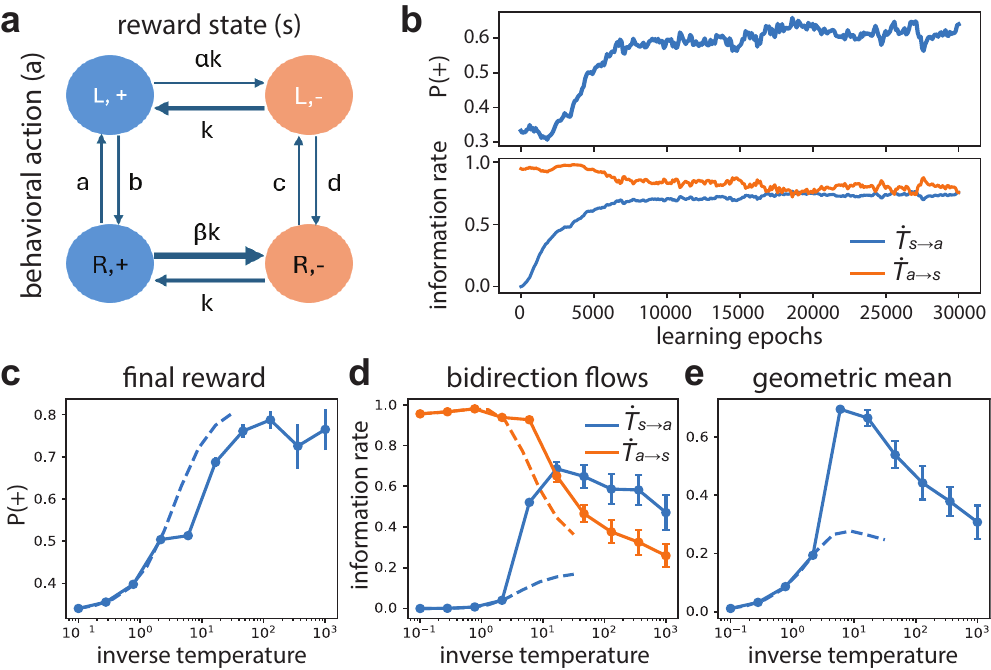}
\par\end{centering}
\caption{Bidirectional information flow reflects constraints in a two-armed bandit task.
\textbf{(a)} A bipartite graph for the two-armed bandit task, with L, R actions and $+$, $-$ rewards. The environment is set by asymmetric rates scaled by $\alpha$ and $\beta$, while control rates a,b,c,d are learned through reinforcement learning.
\textbf{(b)} Reward fraction $P(+)$ and bidirectional information flows across learning epochs. We set $k=1$, $\alpha=0.1$, $\beta=10$, and inverse temperature $\beta_T=10$ for this example.
\textbf{(c)} Steady-state reward reached after learning as a function of the inverse temperature, which controls the reliability of action selection. Other parameters are the same as in (b). Solid lines: numerical results with error bars showing standard deviation from 20 instantiations. Dash lines: analytic results with approximation of local reward at low inverse temperature (Appendix~\ref{appendix: RL model}). 
\textbf{(d)} and \textbf{(e)} Same as (c) but for the bidirectional information flows and their geometric mean, respectively. 
%
\label{fig: RL}}
\end{figure}

The behavioral strategy of biological systems adapt to environmental statistics \cite{brudner2025fly}. A natural question is how the optimal strategy changes across environments, and how these changes are reflected in the information flows. 
We investigate this by analyzing a two-armed bandit problem represented as a four-state bipartite Markov model~(Fig.~\ref{fig: RL}a). The agent can choose between the left (L) and right (R) bandits, while the environment provides either a positive ($+$) or negative ($-$) reward signal. The resulting states are therefore L$+$, L$-$, R$+$, R$-$. The reward dynamics are fixed, whereas the behavioral transition rates between L and R are adapted through reinforcement learning.

To learn these rates, we associate a value function $Q_{RL}(s)$ with each of the four states $s \in \{\text{L}+,\text{L}-,\text{R}+,\text{R}-\}$. At time $t$, the agent occupies state $s_t$ and receives an instantaneous reward $r_t \pm 1$. The state values are updated according to standard Q-learning \cite{sutton1998reinforcement}, 
\begin{equation}\label{eq: Q learning}
    Q_{\rm RL}(s_t) \leftarrow Q_{\rm RL}(s_t) +
\eta \left[ r_t - Q_{\rm RL}(s_t)\right],
\end{equation}
where $\eta$ is the learning rate. The learned values are then mapped to the behavioral transition rates through a softmax (Arrhenius-like) parameterization,
\begin{equation}\label{eq: Q learning rate a}
a=\lambda_0 \exp\left[ 
\beta_T \frac{Q_{\rm RL}(R,+)-Q_{\rm RL}(L,+)}{2}
\right],
\end{equation}
with analogous expressions for the remaining rates. $\beta_T$ is an inverse temperature that controls the reliability of action selection: small $\beta_T$ produces nearly random exploratory behavior, whereas large $\beta_T$ leads to increasingly deterministic exploitation of the higher-valued option. $\lambda_0$ is a baseline rate we set to 1.

As the agent learns while navigating the environment, we measure both the reward and the bidirectional information flows across learning epochs. We find that these quantities relax to their steady-state on the same timescale (Fig.~\ref{fig: RL}b).

As a function of an inverse temperature $\beta_T$ we measure the steady-state reward together with the active and reactive information flows (Fig.~\ref{fig: RL}c-d). At low inverse temperature, the optimal strategy is highly exploratory. In this regime, the reactive information flow is small, while the active information flow is large because the reward remains reliably conditioned on the chosen action. However, at large inverse temperature, the optimal strategy is exploitive, leading to a substantial increase in the reactive information flow and a gradual reduction in the active information flow. 

The increase in the reactive information flow with inverse temperature $\beta_T$ is intuitive, since increasingly deterministic actions become more strongly driven by the reward state. The decrease of active information flow at large $\beta_T$, however, is more  surprising (Fig.~\ref{fig: RL}d). It indicates that optimal performance does not require maximizing active information flow. Instead, there exists an intermediate amount of active information flow that best supports the bandit task.
To further quantify the joint role of the two information flows, we compute the geometric mean of the bidirectional information flows (Fig.~\ref{fig: RL}e). At small inverse temperature, this quantity increases monotonically with the reward. However, this monotonic relationship breaks down as the inverse temperature becomes large. 
We explore the origin of this non-monotonic relationship between information flow and behavioral performance in the following section. 

Together, this simple reinforcement learning example demonstrates how bidirectional information flows characterize distinct reward optimization strategies as the reliability of interactions with the environment changes.


\subsection{Optimal active flow depends on the reliability of behavioral control}

In previous sections, we approximate the contribution of bidirectional information flows at the low information flow limit \eqref{eqn: Phi}. Here we examine the relationship between information flows and navigation performance over a wider range. We revisit the original bipartite graph in Figure \ref{fig:minimal_model}a, which contains two degrees of freedom, $\gamma_u$ and $\gamma_d$, for controlling reaction to the environment. Within the biologically-relevant limit of weak control ($\gamma_d-\gamma_u)\ll1$, we relax the assumption of a low persistence environment ($\alpha\approx 1$) and instead explore the full range of the signal decorrelation factor $\alpha\in[0,1]$.

\begin{figure}
\begin{centering}
\includegraphics[width=1.\columnwidth]{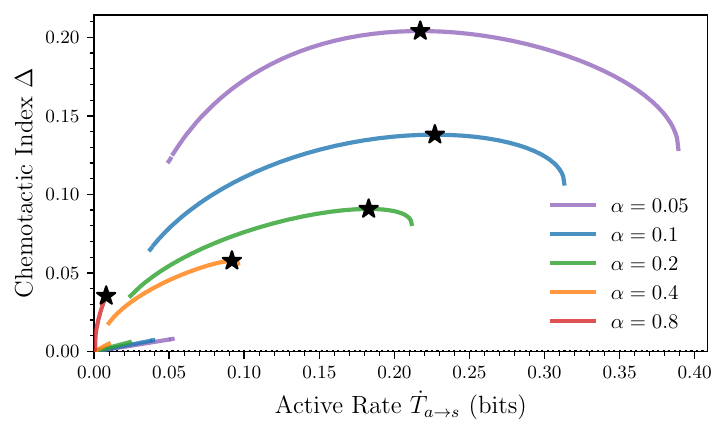}
\par\end{centering}
\caption{Non-monotonic scaling of the chemotactic index, $\Delta$, of the minimal model with the active information flow. For $\lambda =\kappa=1$ and fixed reactive rate $\dot{T}_{s\to a}=0.01$ bits, we optimize $\Delta$ with respect to $\gamma_u$ and $\gamma_d$ at different values of the active information flow $\dot{T}_{a\to s}$. Different colored curves show different values of the signal decorrelation factor $\alpha$. Black stars indicate the maximum performance attainable.
}
\label{fig:nonmonotonicity}
\end{figure}

We fix the rotational diffusion parameter during tumbling $\lambda$ and consider a fixed reactive information flow $\dot{T}_{s\to a}\ll1$. For each value of the environmental noise $\alpha$, we then optimize the remaining rate parameters $\gamma_u$ and $\gamma_d$ to maximize the chemotactic index $\Delta$ at different values of the active information flow $\dot{T}_{a\to s}$. We find that the chemotaxis performance can, in certain regimes, depend non-monotonically on the active information flow (Fig.~\ref{fig:nonmonotonicity}).

Thus, maximizing the active rate $\dot{T}_{a\to s}$ does not always maximize navigation performance. When the 
environment exhibit long correlations 
($\alpha\ll1$), maximum performance is instead attained using an intermediate active information flow. 
This effect is consistent with results from the two-armed bandit task, where the learned policies produce decreased active information flow when the actions are more reliable (Fig.~\ref{fig: RL}d).
In the limit of weakly persistent signals, $\alpha\approx 1$, the performance $\Delta$ instead remains monotonically increasing with the active information flow  $\dot{T}_{a\to s}$. This is precisely the regime in which we derived the metric $\Phi$, and to which the data sets we analyze are most closely related (Fig.~\ref{fig: all data}).

\section{Discussion \label{sec: discussion}}

In this work, we propose an information-theoretic framework to characterize adaptive navigation strategies, based on a minimal model of navigation time series. 
Using transfer entropy rates, we quantify the bidirectional flow of sensory-behavioral information and relate it to navigation performance. The predictions of this minimal framework generalize across experimental measurements (Fig.~\ref{fig: all data}). 
By simultaneously quantifying active and reactive information flows, we capture the diversity of navigation strategies across an isogenic population (Fig.~\ref{fig: bacteria}), the dependency on spatial and environmental conditions (Fig.~\ref{fig: fly}), and distinct learned solutions in a reinforcement-learning task (Fig.~\ref{fig: RL}).
These results show that navigation performance alone does not uniquely determine an agent's strategy. Bidirectional information flow reveals an additional dimension of organization that is not captured by performance metrics alone.
Lastly, we identify regimes of reliable environments in which an intermediate level of active information flow is required for optimal navigation performance (Fig.~\ref{fig:nonmonotonicity}). 
Together, across biological and artificial systems, our results demonstrate that the information loop between sensing and action is fundamental to navigation.\\

Simultaneously with this work, Das et al.\cite{das2026} independently investigated bidirectional information flow during navigation using a complementary theoretical framework. 
They analytically solved continuous-state models with Gaussian noise for agents navigating using spatial and temporal sensory cues and showed that navigation performance is determined by bidirectional transfer entropy, with active and reactive information flows jointly determining performance and exhibiting non-monotonic behavior in regimes of strong feedback. 
In contrast, our work develops a discrete-state minimal model that enables direct comparison with experimental trajectories and provides a framework for analyzing bidirectional information flows from behavioral data across biological systems and artificial agents. 
Together, these complementary studies reach consistent conclusions, highlighting bidirectional information flow as a general organizing principle for navigation.

\paragraph{Comparison to active sensing literature:}

Sensory navigation is a form of active sensing because agents must actively interact with their environment, for example by moving through it, to acquire task-relevant information \cite{kleinfeld2006active, Yang2018-gq, friston2016active, crimaldi2022active, Baker2018-ys, Mobbs2018-au, klyubin2005empowerment}. 
Existing formulations of active sensing often rely on Bayesian inference \cite{Heinonen2023-zy} or optimal control \cite{cellini2024discovering} and typically require the agent to build an internal representation or model of the environment. By contrast, our information-theoretic framework does not explicitly account for an agent's latent internal states. Instead, it estimates information flows directly from the observed behavioral time series. 

In the context of bacterial chemotaxis, previous work has shown that navigation is information-limited in shallow gradients when the feedback from actions to signal is negligible \cite{mattingly2021escherichia}. In steeper gradients, detailed biophysical models demonstrate that feedback between behavior and signal can substantially boost navigation performance \cite{long2017feedback, dufour2014limits}. Our results go beyond previous model-based approaches by taking a data-driven approach. The key advances are an information-theoretic framework that generalizes across scales and a principled approach for estimating information flow directly from experimentally measured navigational trajectories through unbiased state-space representations.
\\

\paragraph{Definition of behavioral states:}

In this work, we coarse-grain continuous behavioral kinematics into discrete behavioral states using an unsupervised clustering approach. Across datasets, we identify stereotyped actions that are consistent with previous characterization, such as run-and-tumble behavioral states in bacteria and steer-and-pirouette dynamics in worms. This approach is inspired by recent work showing that low-dimensional Markov descriptions can emerge from delay-embedded behavioral dynamics \cite{Costa2024-ho}. 
From a theoretical perspective, recent studies have shown how discrete behavioral strategies can arise from task optimization under information constraints \cite{betancourt2026discrete}.
Likewise, statistical approaches have demonstrated that behavior can be decomposed into stereotyped behaviors using latent-variable models \cite{Linderman2019-mj, chen2026state}, and that these discrete latent states can support navigation performance \cite{Verano2023-rp}. 
Although coarse-graining inevitably discards information contained in the raw trajectories, we argue that the dominant behavioral states, which are persistent and stereotyped, provide a useful representation of navigation strategies.

An important assumption of our framework is the uncertainty inherent to both the sensor and behavioral dynamics. In the minimal model (Fig.~\ref{fig:minimal_model}a), both sensory transitions and behavioral actions are treated probabilistically. This assumption reflects the intrinsic variability present in biological behavior and natural environments, and ensures that the information flow rates remain well-defined. 
In the limit when an agent's action has a deterministic effect on the sensory state, the active information flow from behavior to sensor can be infinite. Infinite reactive information flow can similarly arise if the behavior is a deterministic function of the sensory state.
Such deterministic effects are not observed in the experimental systems considered here, where both behavioral variability and environmental fluctuations contribute to stochastic state transitions.
Furthermore, this probabilistic setting allows us to generalize to cases when the environment indeed changes even when the action is fixed, such as fly navigation in complex plumes. 
\\

\paragraph{Extension of navigational model:}

Our minimal model isolates the essential ingredients of sensory navigation---information flow from environment to behavior and the feedback from behavior to the environment. Despite its simplicity, the resulting two-state description captures key features observed in the data, including heterogeneity across trajectories and non-stationary navigation strategies. 
The present framework can be naturally generalized along both behavioral and environmental dimensions. On the behavioral side, latent states could be incorporated to account for memory,  adaptation, hierarchical organization, or planning \cite{Linderman2019-mj, gire2016mice, rigolli2022alternation, friston2016active, berman2016predictability, gosztolai2020cellular}. For example, one could introduce intermediate memory states in a graphical model, analogous to the information bottleneck method, to describe how sensorimotor information is processed. On the environmental side, one can move beyond simple low-dimensional sensory landscapes and consider more complex environments in which exploration-exploitation tradeoffs become salient \cite{Vergassola2007-yb, Celani2014-ki, Still2012-vx, biswas2023mode, klyubin2005empowerment}. \\

Navigation is a universal computational problem tackled by systems ranging from single cells to nervous systems and artificial agents. 
Bidirectional information flow provides a quantitative description of the behavior–environment information loop and reveals diverse adaptive navigation strategies that are not captured by performance metrics alone.
The optimized solutions of the minimal model further demonstrate that different navigational regimes require distinct balances between active and reactive information flows.
Looking forward, we expect this framework to help constrain candidate biological mechanisms underlying navigation and guide the development of learning algorithms for adaptive agent–environment interactions.\\


\begin{acknowledgments}
We are indebted to Michael Abbott for initial ideas about parameterizing the bipartite model, and to Marianne Bauer for pointing us to the two-armed bandit task. We thank Ben Machta, Henry Mattingly, Jose Betancourt, Alexandra Walczak, and Pieter Rein ten Wolde for feedback on the project. 
This work was supported by NIH awards R35GM158058 (KSC, TE) and RF1NS132840 (KSC, DAC, TE), by the Alfred P. Sloan Foundation Award G-2023-19668 (KSC, TE), by Mossman and NSERC Postdoctoral Fellowships (MPL), and by the Kavli Postdoctoral Fellowship (KSC).
\end{acknowledgments}

\appendix

\section{Steady-state solution of the minimal model \label{appendix: steady-state}}

Solving $\pi Q=0$ for the steady-state probabilities $\pi=(\pi_{Ru}, \pi_{Rd}, \pi_{Tu}, \pi_{Td})$ we get:
\begin{equation}
\pi
=
\frac{1}{Z}
\begin{pmatrix}\label{eq:pi_steady_state}
\alpha\left(2\lambda+1\right)+\gamma_d \\
\alpha\left(2\lambda+1\right)+\gamma_u \\
\alpha\lambda\left(\gamma_d+\gamma_u\right)+\gamma_u\left(\alpha+\gamma_d\right)
\\
\alpha\lambda\left(\gamma_d+\gamma_u\right)+\gamma_d\left(\alpha+\gamma_u\right)
\end{pmatrix}^T
\end{equation}
with the normalization factor 
\begin{equation}\label{eq:pi_steady_Z}
    Z = \alpha\left(\gamma_d+\gamma_u\right)(2 \lambda+1)+2 \alpha(2 \lambda+1)+2 \gamma_d \gamma_u+\gamma_d+\gamma_u
\end{equation}

\section{Single versus multi-step transfer entropy rates
\label{appendix: single vs multi-step}
}

The transfer entropy rate
\begin{equation}\label{eq:transfer_entropy_rate_full}
    \dot{T}_{s \rightarrow a} = \lim_{dt \rightarrow 0}  
     \frac{1}{dt}\left[H(a_{t+dt}|a_{:t}) - H(a_{t+dt}|a_{:t},s_{:t})\right]
\end{equation}
quantifies how much knowing the signal time series, $s_{:t}$, up to time step $t$ reduces the uncertainty about the next action to take after time step $dt$, $a_{t+dt}$, given the history of past actions $a_{:t}$. Here $H$ is the Shannon entropy, and the subscripts $t$, $t+dt$ and $:t$ represent the current time $t$, the next time $t+dt$, and the full history up to the current time \cite{Schreiber2000-ir, mattingly2021escherichia}. $\dot{T}_{a \rightarrow s}$ is defined similarly. 

In this work we used instead the simpler one-step transfer entropy rates that only depends on the previous step~\eqref{eq:transfer_entropy_rate_one_step}. For a continuous Markov process the one-step transfer entropy rates~\eqref{eq:transfer_entropy_rate_one_step} become:
\begin{equation}\label{eq:transfer_entropy_rate_one_step_markov}
    \dot{T}_{s \rightarrow a}=\sum_{s, a} \pi_{s, a} \sum_{a^{\prime} \neq a} q_{a \rightarrow a^{\prime} \mid s} \ln \frac{q_{a \rightarrow a^{\prime} \mid s}}{\bar{q}_{a \rightarrow a^{\prime}}}
\end{equation}
where $\pi_{s, a}$ is the steady-state probability of the joint state $\{s,a\}$, $q_{a \rightarrow a^{\prime} \mid s}$ is the transition rate from $a$ to $a^{\prime}$ conditional on $s$, and the marginalized transition rate is $\bar{q}_{a \rightarrow a^{\prime}}=\sum_s q_{a \rightarrow a^{\prime} \mid s} \pi(s \mid a)$ with $\pi(s \mid a)$ the steady-state conditional probability of $s$ given $a$. $\dot{T}_{a \rightarrow s}$ is defined similarly. Plugging in the steady-state probabilities \eqref{eq:pi_steady_state} and the conditional transition rates \eqref{eq:Qmatrix} we obtain the one-step transfer entropy rates for the minimal model~\eqref{eq:entropy_transfer_rate_s_a}-\eqref{eq:entropy_transfer_rate_a_s}.

While we have focused in this Article on single-step transfer entropy rates, defined as
\begin{subequations}
\begin{align}
\dot{T}_{s\to a} & \equiv \lim_{\mathrm{d}t\to0} \frac{1}{\mathrm{d}t}\left[H(a_{t+dt}|a_{t}) - H(a_{t+dt}|a_{t},s_{t})\right],\\
\dot{T}_{a\to s} & \equiv \lim_{\mathrm{d}t\to0} \frac{1}{\mathrm{d}t}\left[H(s_{t+dt}|s_{t}) - H(s_{t+dt}|s_{t},a_{t})\right],
\end{align}
\label{eq:single_step_TE}
\end{subequations}
our main results still hold qualitatively, and in many cases quantitatively, for the full-history transfer entropy rates
\begin{subequations}
\begin{align}
\dot{T}_{s\to a}^\mathrm{(hist)} & \equiv \lim_{\mathrm{d}t\to0} \frac{1}{\mathrm{d}t}\left[H(a_{t+dt}|a_{:t}) - H(a_{t+dt}|a_{:t},s_{:t})\right[,\\
\dot{T}_{a\to s}^\mathrm{(hist)} & \equiv \lim_{\mathrm{d}t\to0} \frac{1}{\mathrm{d}t}\left[H(s_{t+dt}|s_{:t}) - H(s_{t+dt}|s_{:t},a_{:t})\right].
\end{align}
\end{subequations}
To show this, in the SI we derive semi-analytic expressions for the full-history transfer entropy rates, which allow us to efficiently compute them numerically. Supplementary Figure \ref{SI:TE_Ratio} shows the ratio between the single-step and full-history transfer entropy rates across a wide parameter range. The reactive rates are comparable in magnitude throughout the range, while the active rates differ significantly only in the limit of highly persistent environment and strong control. Most importantly, the rates differ by less than $10\%$ in the regime in which we derived the metric $\Phi$, corresponding to $1-\alpha\ll1$ and $(\gamma_d-\gamma_u)\ll1$.

\begin{suppfigure}[!h]
\centering
\includegraphics[width=\columnwidth]{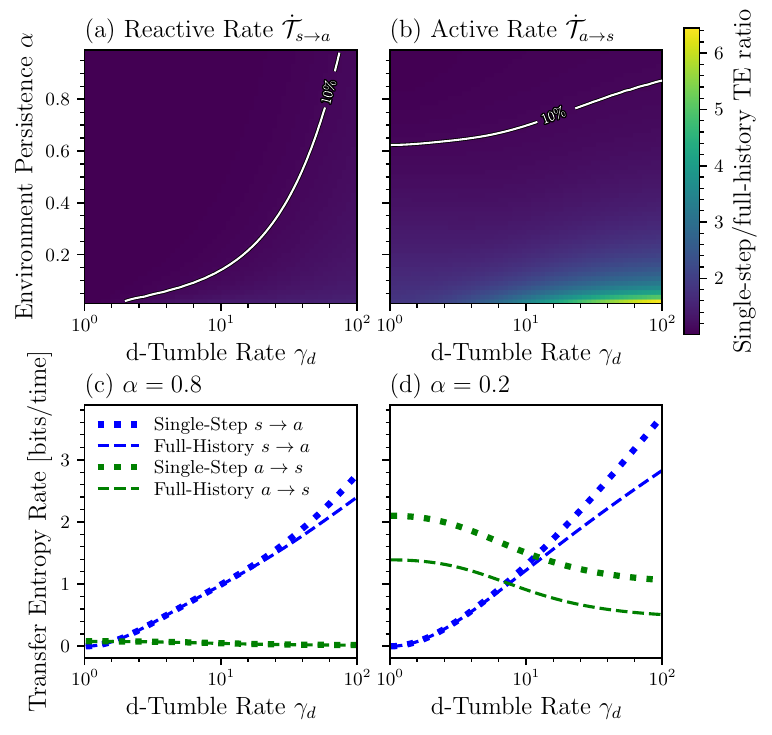}
\caption{Heatmaps showing the ratio between single-step and full-history transfer entropy rates for both the reactive (a) and active (b) rates, as functions of $\gamma_d$ and $\alpha$. We take $\kappa=1$, $\lambda=10$, and $\gamma_u = \kappa/\gamma_d$. (c) and (d) Cross-sections showing the single-step and full-history transfer entropy rates as functions of $\gamma_d$ for $\alpha=0.8$ (c) and $\alpha=0.2$ (d). All transfer entropy rates shown are in bits.}
  \label{SI:TE_Ratio}
\end{suppfigure}



\section{Exploring other navigational metrics
\label{appendix: metrics}
}


Given the state transitions of behavioral action and sensory signals, we explore other possible metrics to compare with our proposed bidirectional flow metric $\Phi$. In addition to specific terms in $\Phi$ (Table \ref{tab:model_comparison}), we report in the main text $R^2$ values for the following performance predictors: the mutual information between two time series $I(s,a)$, the transfer entropy from sensing to action $\dot{T}_{s \rightarrow a}$, and the entropy production rate $\dot{S}$ in the bipartite graph (Figure~\ref{fig:minimal_model}i). Here we provide the equations for these metrics.

The mutual information between sensing and action follows:
\begin{equation}
    I(a , s)=\sum_{a, s} \pi_{a, s} \log \frac{\pi_{a, s}}{\pi_a \pi_s}
\end{equation}
where $\pi_{ij}$ is the steady-state distribution of states on the bipartite graph.

The non-equilibrium flux on the bipartite graph follows:
\begin{equation}
    J_{i \rightarrow j}=\pi_i k_{i \rightarrow j}-\pi_j k_{j \rightarrow i}
\end{equation}
where $k$ are the rate parameters along the edges of the graph. The scalar value is computed by summing over all directions of $i\rightarrow j$ to capture the net probability current.

The entropy production rate $\dot{S}$ follows:
\begin{equation}
    \dot{S} = \sum_{i \neq j} \pi_i Q_{i j} \ln \left(\frac{\pi_i Q_{i j}}{\pi_j Q_{j i}}\right)
\end{equation}
where $Q$ is the system rate matrix and $\pi$ is the steady-state distribution.

\begin{table}[t]
\caption{
Model comparison of the information theoretic predictor.
Positive $\Delta$BIC (Bayesian Information Criterion) indicate support for the full predictor $\Phi$. 
Number of the observed navigation trajectory ensembles is $N$. We compare predictors with (w) and without (wo, only $\sqrt{\dot{T}_{s\rightarrow a}}$) the geometric mean term $\sqrt{\dot{T}_{s\rightarrow a}\dot{T}_{s\rightarrow a}}$.
RMSE improvement is defined as the percentile increase
$100\left(1-\mathrm{RMSE}_{\rm w}/\mathrm{RMSE}_{\rm wo}\right)$.
}
\label{tab:model_comparison}
\centering
\begin{tabular}{lrrr}
\hline
Dataset & N & $\Delta$BIC & RMSE (\%) \\
\hline
Minimal Markov model  & 16000 & 4452.31 & 75.1 \\
Chemotaxis model    & 200  & 301.57  & 52.9 \\
RL agent      & 461  & 180.82  & 17.8 \\
Bacteria experiments & 70   & 18.08   & 12.1 \\
Worm experiments    & 121   & 26.80   & 10.5 \\
Fly experiments     & 104    & 19.01    & 8.7 \\
\hline
\end{tabular}
    \label{table:model comparison}
\end{table}

\section{Defining state-space from continuous navigational trajectories
\label{appendix: state-space}
}

We process navigational measurements from bacteria, worms, and flies. For bacteria, the \emph{E. coli} (RP437) cells are tracked in gradient of methyl-aspartate through methods previously described in \cite{Waite2018-lu}, but using a gradient length scale of 0.4 $\mu$m/mm instead of 0.1. For worms, animals (N2 strains) are tracked on an agar plate with controlled butanone odor landscape through methods described in \cite{Chen2023-fy}. For flies, animals are tracked in complex or straight smoke plumes through methods described in \cite{Demir2020-rk}.

We coarse grain the navigational kinematics through a method inspired by maximum predictive modeling of animal posture dynamics \cite{Costa2024-ho}. The data structure are ensembles of measured navigational trajectories across time $T$ and with kinematic dimension $d$ (for instance, $d=2$ when we consider translational and rotational speed). We embed these kinematics with a time window $K$, forming trajectories in a $K\times d$ dimension space, the cluster them into $N=2$ classes. As a function of window size of $K$, we observe that the steady-state entropy rate drops drastically at a certain length. This length is then chosen to be the embedding window to define behavioral actions (Supplementary Figure~\ref{SI:Markov}).

\begin{suppfigure}[!h]
\centering
\includegraphics[width=.66\columnwidth]{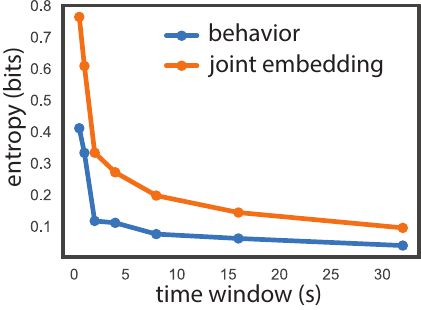}
\caption{Time delayed embedding to find consistent state transition models. As a function of the time window size, we build two-state Markov models fitted to data and computed the steady-state entropy rate. Models for behavior-only and joint embedding with sensory signal are shown. This example data is from bacteria chemotaxis data, showing $\sim 2$ s embedding gives a consistent state-space description.
}
  \label{SI:Markov}
\end{suppfigure}

The behavioral kinematics and signal from the sensory environment are often continuous and have complex dynamics. To simplify the calculation, we construct maximally predictive Markov states from continuous time series, following ideas from recent work \cite{Costa2024-ho}. This approach defines states in the delay embedding space and then approximates the transfer operation with a discrete Markov model. With the number of states $N$, time window for embedding $h$, and discrete time step $\delta t$, the short-term entropy rate follows:
\begin{equation}
    h_{\delta t}(N, h)=-\frac{1}{\delta t} \sum_{i, j}^N \pi_i P_{i j} \log P_{i j}
\end{equation}
where $\pi$ is the steady state, $P_{ij}$ is the Markov transition matrix, and the states are in the $K$-step time delay embedded space. We scan this entropy rate as a function of window size $K$ and use time step $\delta t$ from the experimental sampling rate.

Given the discrete time series of binary behavioral states $a_t$, we apply the same-length binary time series for the sensory signal $s_t$. 
We rank order trajectories according to the fraction of time in a given state. This corresponds to the tumble bias in bacteria, fraction of pirouetting in worms, and fraction of stops in flies. The ordered trajectories are binned into 100 groups, each with ensembles of 20-50 trajectories to increasing the number of observed state transitions. We remove the lowest 10-20 ensembles due to their limited numbers of observed behavioral state transitions. For each ensemble of pooled trajectories, we then compute the transfer entropy rates directly using equations \eqref{eq:single_step_TE}, where the conditional entropy terms are directly computed from the empirical binary states. The time step $dt$ is directly taken from the sampling rates of each data set: 20 Hz for bacteria, 2.8 Hz for worms, and 60 Hz for flies. With base-2 logarithm in the entropy calculation, the estimates have units of bits per second. 
With finite trajectories from experimental measurements and simulations, we apply finite-size correction described in the following section.

\section{Finite size and coarse grained effects for information estimation
\label{appendix: finite-size}
}

When applying information-theoretic measures to data, at least two sources of bias must be considered: finite-sample bias and biases introduced by discretization.
We correct for the finite-data bias by scaling the measurements with data length and extrapolating for the infinite data limit \cite{strong1998entropy}. Across measurements, we can infer non-zero information flow through this method (Supplementary Figure~\ref{SI:finite}). 
Given finite data, we subsample the data into 4–5 fractions, independently resample each fraction 10–20 times, and fit the estimated information rate as a quadratic function of the inverse sample size. The extrapolated value at infinite sample size is used as the final estimate.

\begin{suppfigure}[!h]
\centering
\includegraphics[width=.7\columnwidth]{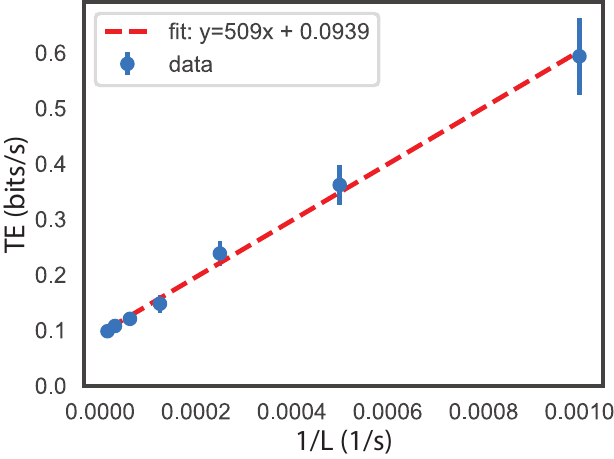}
\caption{Finite size correction for information estimation. We follow previous methods \cite{strong1998entropy} to correct for finite size over-estimation of information content by extrapolating non-zero transfer entropy (TE) rate at infinite data length ($L$). Here the extrapolated TE rate is $\approx 0.09$ bits/s. This example is for bacteria chemotaxis, with error bars showing 100 repeated subsampling.
}
  \label{SI:finite}
\end{suppfigure}

Another approximation we made was to discretize behavioral time series into finite states. We confirm the consistency of these state estimates through delay embedding. However, following information inequality, this coarse-graining procedure will likely underestimate, rendering the reported information flow here a lower-bound of the true system \cite{kraskov2004estimating}. One can formalize better estimation with tighter bounds for the information flow in future work.

\section{RL models for navigation and two-armed bandit task}\label{appendix: RL model}
For agent-based navigational models, we model bacteria chemotaxis following a simplified model reported previously \cite{long2017feedback}. The RL agent is a discrete time model moving along a one-dimensional environment. Similar to the minimal model, the actions $a\in \{R,T\}$ correspond to a high probability to continue the correct motion (with $\epsilon$ small probability of flipping) and $s \in \{u,d\}$ correspond to the up or down-gradient states. The reward is $+1$ for climbing up and $-1$ for down the gradient. The policy $\pi(a|s)$ is a logistic function with two parameters, and we apply the policy gradient method to optimize for the expected total reward \cite{sutton1998reinforcement, Verano2023-rp}.

For the two-armed bandit task, we associate a value table $Q_{\rm RL}(S)$, where each state $s \in \{ L+, L-, R+, R-\}$. By running stochastic stimulation, an agent will receive reward time series $r_t$. We update the values with standard Q-learning \eqref{eq: Q learning}.
The learned state values mapped to action-switching rates through a softmax-like parameterization \eqref{eq: Q learning rate a}.

At low inverse temperature $\beta_T$, the Q-values can be estimated through ``local'' expected reward without running the iterative learning in \eqref{eq: Q learning}. This is done by approximating the expected rate to transition to reward given a state. 
For instance, $Q_{\rm RL}(L,+) \approx \frac{a}{a+\alpha k}$ reflects the expected transition to continue with reward $+$ from this state. Given these approximations, we then solve the self-consistent equations to find the corresponding rate parameters. As shown in Figure \ref{fig: RL}, this short-term local calculation for reward is good approximation before it diverges at higher $\beta_T$.

\section{Data and code}

Data and code used to generate figures in this work are available: \url{https://github.com/emonetlab/navigation-info-loop}\\

\newpage

\bibliography{info_flow_navigation}
\clearpage
\onecolumngrid
\newcommand{\includesipage}[1]{%
  \thispagestyle{empty}%
  \noindent\makebox[\textwidth][c]{%
    \includegraphics[page=#1,width=\paperwidth,height=\paperheight,keepaspectratio]{SI.pdf}}%
  \newpage}
\includesipage{1}
\includesipage{2}
\includesipage{3}
\includesipage{4}
\includesipage{5}
\includesipage{6}
\end{document}